# Enhanced Prediction of Three-dimensional Finite Iced Wing Separated Flow Near Stall

Maochao Xiao (肖茂超)[*], Yufei Zhang (张宇飞)[†]

*Tsinghua University, 100084 Beijing, People's Republic of China*

Feng Zhou (周峰)[‡]

*Shanghai Aircraft Design and Research Institute, 201210 Shanghai, People's Republic of China*

## Abstract

Icing on three-dimensional wings causes severe flow separation near stall. Standard improved delayed detached eddy simulation (IDDES) is unable to correctly predict the separating-reattaching flow due to its inability to accurately resolve the Kelvin-Helmholtz instability. In this study, a shear layer adapted subgrid length scale is applied to enhance the IDDES prediction of the flow around a finite NACA (National Advisory Committee for Aeronautics) 0012 wing with leading edge horn ice. It is found that applying the new length scale contributes to a more accurate prediction of the separated shear layer (SSL). The reattachment occurs earlier as one moves towards either end of the wing due to the downwash effect of the wing tip vortex or the influence of end-wall flow. Consequently, the computed surface pressure distributions agree well with the experimental measurements. In contrast, standard IDDES severely elongates surface pressure plateaus. For instantaneous flow, the new length scale helps correctly resolve the rollup and subsequent pairing of vortical structures due to its small values in the initial SSL. The computed Strouhal numbers of vortical motions are approximately 0.2 in the initial SSL based on the vorticity thickness and 0.1 around the reattachment based on the separation bubble length. Both frequencies increase when moving towards the wing tip due to the downwash effect of the tip vortex. In comparison, the excessive eddy viscosity levels from the standard IDDES severely delay the rollup of spanwise structures and give rise to "overcoherent" structures.

[*] Graduate student, School of Aerospace Engineering.
[†] Associate professor, School of Aerospace Engineering; corresponding author, email: zhangyufei@tsinghua.edu.cn.
[‡] Research scientist

# I. Introduction

Leading edge ice accretions tend to cause severe flow separation for aircraft, and the essentially unsteady flow leads to strong fluctuating aerodynamic loads, which greatly deteriorate aircraft performance [1]. The iced airfoil flow with leading edge horn ice is dominated by an ice-induced unsteady separating-reattaching shear layer when the angle of attack is not overly high [1]. Before 2000, only Reynolds-averaged Navier–Stokes (RANS) simulations were conducted, which are neither accurate in predicting mean aerodynamic loads nor capable of reproducing the flow unsteadiness. They typically underestimate lift coefficients by producing elongated ice-induced separation bubbles, which is to a great extent caused by the incapability of RANS methods in correctly modelling the nonequilibrium turbulence in the ice-induced separated shear layer (SSL). The weakness becomes increasingly prominent as the stall condition is approached [2]. Afterwards, attention gradually shifted to hybrid RANS-large eddy simulation (RANS-LES) methods, of which detached eddy simulation (DES) and its variants have gained the spotlight [3]. However, existing numerical studies of iced lifting surfaces show a lack of [3]:

(1) methodologies that predict the flow fields accurately and robustly;

(2) numerical investigations of three-dimensional (3D) iced wings.

The laminar-to-turbulent transition commonly occurs in ice-induced SSLs [4]. However, DES-type methods with subgrid length scale evaluated as the maximum cell edge length are incapable of correctly resolving flow transitions because they generate excessive eddy viscosity, which is especially true when the grid cells are highly anisotropic [5-7]. Consequently, the rollup of two-dimensional (2D) structures is retarded, which in turn delays flow reattachment. It has been shown by the current authors [8] that a shear-layer adapted subgrid length scale helps enhance the improved detached eddy simulation (IDDES) of 2D iced airfoil flows. However, it is still unknown whether it helps 3D iced wing flows.

Although a number of high-resolution studies have been conducted for 2D iced airfoils, there have been few investigations on 3D iced wings. Table 1 lists hybrid RANS-LES simulations of iced airfoil/wing flows from 2015. Based on the authors' knowledge, iced wings have only been studied by Stebbins et al. [9]. In their study, an 8.9%-scaled CRM65 wing with leading edge ice was simulated via IDDES [9]. Although the method produces reasonable lift and drag, it fails to accurately predict force moments since the predicted leading edge separation shows remarkable discrepancies.

The current study aims to improve the IDDES prediction of 3D iced wing separated flows near stall by using a shear layer adapted subgrid length scale. Additionally, the effects of the wing-tip vortex are analysed for a complete understanding of iced wing flows.

Table 1 Hybrid RANS-LES simulations of iced lifting surfaces with leading edge ice

| Reference | Base airfoil | Airfoil/wing | Ice shape | Length of span ($L_z/c$) | Number of grid nodes (million) | Turbulence model |
|---|---|---|---|---|---|---|
| Alam et al. [10] | GLC 305 | Airfoil | Ice 944 | 0.5 | 15 | DDES, DHRL |
| Zhang et al. [11] | GLC 305 | Airfoil | Ice 623 | 0.3 | 9.5 | Zonal DES |
| Butler et al. [12] | NACA 0012 | Airfoil | 3.5-minute ice | \ | 14 | IDDES |
| Stebbins et al. [9] | 65% CRM | Wing | Simplified ice | 3.6 | 28 | IDDES |
| Xiao et al. [13] | NLF 0414 | Airfoil | Ice 623 | 0.4 | 16 | Wall-modelled LES |
| Molina et al. [14] | GLC 305 | Airfoil | Ice 944 | 0.5 | 10 | DDES |
| Xiao et al. [15] | 30P30N | Airfoil | 6-minute ice | 0.1 | 61 | Wall-modelled LES |
| Lee et al. [16] | 30P30N | Airfoil | 6-minute ice | 0.04 | 46 | Coarse-grid LES |
| Xiao et al. [8] | GLC 305 | Airfoil | Ice 944 | 0.4 | 30 | IDDES |
| Zhang et al. [17] | GLC 305 | Airfoil | Ice 944 | 0.5 | 88 | IDDES |

DHRL: dynamic hybrid RANS/LES

# II. Methodology

**A. IDDES Formulation and Shear Layer Adapted Subgrid Length Scale**

The IDDES based on the SST $k-\omega$ turbulence model is obtained by replacing the RANS length scale with a hybrid length scale in the turbulent kinetic energy (TKE) transport equation [18]:

$$\frac{\partial(\rho k)}{\partial t}+\frac{\partial(\rho u_j k)}{\partial x_j}=P_k-\frac{\rho k^{\frac{3}{2}}}{l_{\text{hybrid}}}+\frac{\partial}{\partial x_j}\left[(\mu+\sigma_k\mu_t)\frac{\partial k}{\partial x_j}\right] \quad (1)$$

The hybrid length scale $l_{\text{hybrid}}$ is defined as

$$l_{\text{hybrid}}=f_d(1+f_e)l_{RANS}+(1-f_d)l_{LES} \quad (2)$$

where $f_d$ is a switching function; it is 1.0 for RANS mode and 0.0 for LES mode. The RANS length scale is $l_{RANS}=\sqrt{k}/\omega$, and the LES length scale is

$$l_{LES}=C_{DES}\cdot\min(\Delta_{\text{wall}},\Delta_{\text{free}}) \quad (3)$$

where $\Delta_{\text{wall}}$ is activated in the proximity of walls to mitigate the log-layer mismatch issue; $\Delta_{\text{free}}$ works away from walls and is classically defined as the maximum cell edge length $\Delta_{\text{max}}$. However, such a definition is too conservative for free shear flows to unlock Kelvin-Helmholtz (K-H) instability, especially when the grid cells are strongly anisotropic [5]. To remedy this issue, a shear layer adapted subgrid length scale is proposed [19]:

$$\Delta_{SLA}=\Delta_\omega F_{KH}(\text{VTM}) \quad (4)$$

$$\Delta_\omega=\frac{1}{\sqrt{3}}\max_{n,m=1,\ldots,8}|\mathbf{n}_\omega\times\mathbf{r}_{mn}| \quad (5)$$

where $\mathbf{n}_\omega$ denotes the unit vorticity vector and $\mathbf{r}_{mn}$ is the vector formed by two arbitrary vertices of a cell. The vorticity-related length scale $\Delta_\omega$ can be regarded as a maximum cell edge length definition in the plane normal to the local vorticity vector. It helps reduce eddy viscosity if the lateral cell size is the largest dimension (e.g., pencil- and ribbon-shaped cells) in planar shear layers. The vortex titling measurement VTM evaluates how much a strain tensor tilts a vorticity vector towards another direction. It is zero for planar shear layers, which makes the scaling function $F_{KH}(\text{VTM})$ very small (0.01 herein) to further reduce eddy viscosity levels. More details about the length scale and its implementation into SST-IDDES are provided in [19-21]. Hereafter, the IDDES with $\Delta_{\text{free}} = \Delta_{\max}$ and $\Delta_{\text{free}} = \Delta_{\text{SLA}}$ will be referred to as IDDES and IDDES-SLA, respectively.

**B. Numerical Setup**

The Navier–Stokes equations are solved using the in-house structured finite volume solver NSAWET [8, 22]. The inviscid flux is discretized via a hybrid central/upwind scheme; the central part is a fourth-order central difference scheme, and the upwind part is a fifth-order Roe/WENO scheme. The blending factor is calculated based on the ratio $l_{\text{hybrid}}/l_{\text{RANS}}$ [23]:

$$F_{\text{inviscid}} = (1-\sigma)F_{\text{central}} + \sigma F_{\text{upwind}} \tag{6}$$

$$\sigma = \max\left(\tanh\left(\frac{C_3}{1-C_4}\max\left(\frac{l_{\text{hybrid}}}{l_{\text{RANS}}} - C_4, 0\right)\right), \sigma_{\min}\right) \tag{7}$$

where $C_3 = 4.0$, $C_4 = 0.6$ and $\sigma_{\min} = 0.1$. In LES-resolved regions, $l_{\text{hybrid}} = l_{\text{LES}}$ and $l_{\text{LES}}/l_{\text{RANS}} < 0.6$ are common such that $\sigma = 0.1$. In RANS-modelled boundary layers, $l_{\text{hybrid}} = l_{\text{RANS}}$ so that $\sigma$ is approximately 1.0. The viscous flux is approximated via a second-order central difference scheme. Time advancement is performed by an implicit dual-time-step lower-upper symmetric Gauss-Seidel (LU-SGS) scheme.

# III. Computational Configuration and Mesh

The computational model is a NACA 0012 semispan unswept wing with leading edge horn ice [24]. The wing has a chord length $c$=0.38 m and semispan $b$=2.5$c$. The ice is the simplification of a measured ice shape accreted in the NASA Icing Research Tunnel, with the roughness excluded since aerodynamic effects are negligible for iced wings with leading edge horn ice [1]. Figure 1 (a) shows a cross-section of ice accretion. It is characterized by a height $H/c$=3.2%, angle $\theta$=35 deg and location measured by the surface length $s/c$=1.0%. In the aerodynamic experiment conducted at Ohio State University, the model is installed in a wind tunnel with a test section of 1.01 m by 1.40 m (Figure 1 (b)). In the current simulation, tunnel walls are included to incorporate their blockage effects [24, 25]. The inflow condition is set as the total pressure and temperature, and the outflow condition is the static pressure. The sidewalls are set with adiabatic nonslip conditions, but the wall where the model is installed is set as a slip wall from the inlet to 0.3$c$ upstream of the model leading edge to simulate the suction in the experiment. The Reynolds number based on the chord length and inflow velocity is $1.5 \times 10^6$, the Mach number is 0.2, and the angle of attack (AoA) is 8 deg. This AoA corresponds approximately to the stall condition. The experimental aerodynamic results are provided by Bragg et al. [24, 25]. The current iced unswept wing is simple but indeed representative since the flow exhibits not only extended ice-induced separation but also the effects of the tip vortex and end-wall interactions.

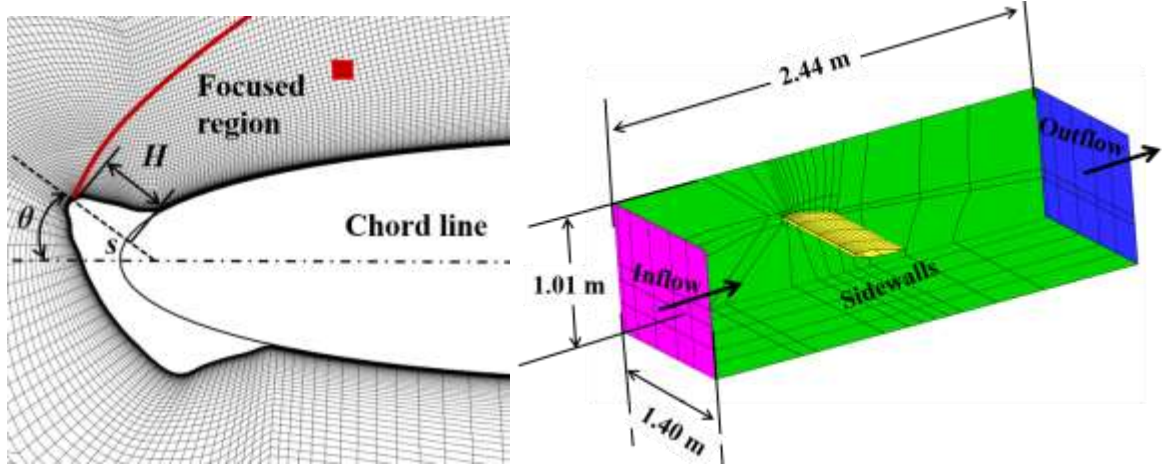

(a) Computational grid around ice accretion  (b) Computational domain

Figure 1 Computational grid and domain.

Table 2 describes the two sets of grids used in the study. $N_z$ denotes the cell number in the spanwise direction. The cells are clustered around the ice as well as its downstream region. For the fine grid, the ice is surrounded by 80 cells circumferentially, and the cell size is $\Delta x \approx \Delta y \approx 0.078$ near the red marker shown in Figure 1 (a). The spanwise cell size is $\Delta z/H=0.25$ for most of the span except close to the wing tip and root. The total number of cells is 17.14 million. The coarse grid is obtained by coarsening the fine grid by approximately 0.85 in each direction. It is worth noting that the current spanwise cell sizes are much larger than those in 2D iced airfoils because the current span ($2.5c$) is 5-8 times those ($0.3$-$0.5c$) for 2D iced airfoils, but the total number of cells is approximately the same as in most 2D iced airfoil studies (Table 1). This unavoidable grid anisotropy makes an accurate prediction even more challenging.

Table 2 Grid details

| Grid | $N_z$ | $N_{x,\text{ice}}$ | $(\Delta x/H)_{\text{focus}}$ | $\Delta z/H$ | $\Delta y_w/H$ | $N_{\text{total}}$ (million) |
|---|---|---|---|---|---|---|
| Fine | 436 | 80 | 0.078 | 0.25 | $6.25 \times 10^{-4}$ | 17.14 |
| Coarse | 376 | 64 | 0.091 | 0.29 | $6.25 \times 10^{-4}$ | 9.98 |

# IV. Computational Results

The unsteady simulations are initialized by converged RANS solutions. The physical time step is $\Delta t U_\infty / c = 0.001$, ensuring the CFL number $\Delta t U_\infty / \Delta x \approx 1$ along the SSL. The number of subiterations is 40 for a second-order temporal accuracy. The last 25 time units ($c/U_\infty$) are collected for statistical analysis after the simulations become fully developed.

## A. Statistical Results

The temporal variations in the integrated forces and pitching moment on the fine grid are shown in Figure 2. The moving average is performed to ensure that the time average has reached convergence. Table 3 lists the statistical results, in which the relative difference is the percentage by which an IDDES value is greater than its IDDES-SLA counterpart on the fine grid. Although both methods produce similar lift coefficients, IDDES predicts a nose-down moment and higher drag. Additionally, IDDES yields stronger fluctuations, which will be further discussed below. The results obtained via IDDES-SLA on the coarse grid are also given in Table 3. The time-averaged forces and moment are almost identical to the results on the fine grid, although the instantaneous values exhibit higher levels of fluctuations.

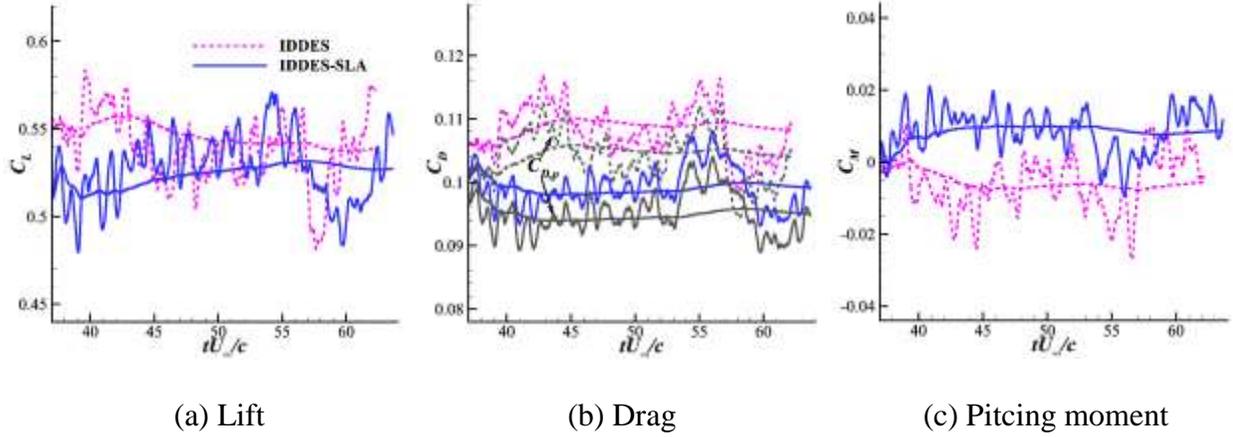

(a) Lift           (b) Drag           (c) Pitcing moment

Figure 2 Integrated aerodynamic force (moment) coefficients on the fine grid.

Table 3 Integrated lift, drag and pitching moment

| Method | $C_L$ | $C_{L,\,RMS}$ | $C_D$ | $C_{D,\,RMS}$ | $C_M$ | $C_{M,\,RMS}$ |
| --- | --- | --- | --- | --- | --- | --- |
| IDDES | 0.537 | 0.0195 | 0.108 | 0.0040 | -0.0059 | 0.0077 |
| IDDES-SLA | 0.527 | 0.0182 | 0.099 | 0.0033 | 0.0085 | 0.0060 |
| IDDES-SLA, coarse | 0.538 | 0.0215 | 0.100 | 0.0044 | 0.0081 | 0.0082 |
| Relative difference | 1.9% | 7.1% | 9.0% | 21.2% | -169.4% | 28.33% |

Figure 3 shows the sectional forces and pitching moment. The results predicted via IDDES-SLA on both grids are very close, except slightly higher lift values are produced on the coarse grid between $z/b$=0.2-0.4. On the fine grid, IDDES-SLA predicts lift coefficients that are more consistent with the experiment [24], while IDDES yields higher lift values, and the discrepancy reaches up to 5% near the midspan. In regard to the drag and pitching moment, the distinctions between the two methods are amplified. The IDDES-predicted drag is approximately 15% higher than that of IDDES-SLA near the root. Figure 3 (a) and (b) integrate only the pressure force; the frictional force is not involved. Figure 3 (c) further compares the calculated $C_L$-$C_D$ of the midspan section with the experimental sectional drag polars for the 3D iced wing and 2D iced airfoil [24]. The computational frictional drag is estimated as 4% pressure drag because Figure 2 (b) indicates that the pressure drag $C_{D,p}$ accounts for approximately 96% of the total drag. It can be seen that the drag values predicted via IDDES-SLA on both grids are almost identical and agree well with the experiment [24], while the IDDES result is 13% higher. What is worse is that IDDES predicts nose-down moments for the inner 70% span, while IDDES-SLA yields nose-up moments except in the wing tip region. Hence, the current results confirm the review of Stebbins et al. [3] that drag and moment are more challenging to compute. In the following, surface and spatial flows are scrutinized to further demonstrate the distinctions of the two methods.

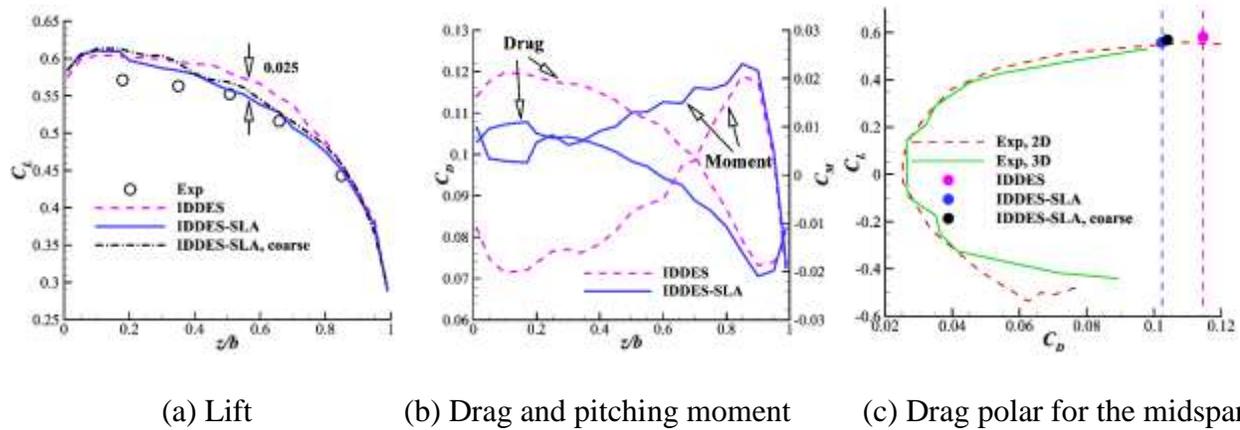

(a) Lift      (b) Drag and pitching moment      (c) Drag polar for the midspan

Figure 3 Sectional forces and pitching moment. Experimental results [24].

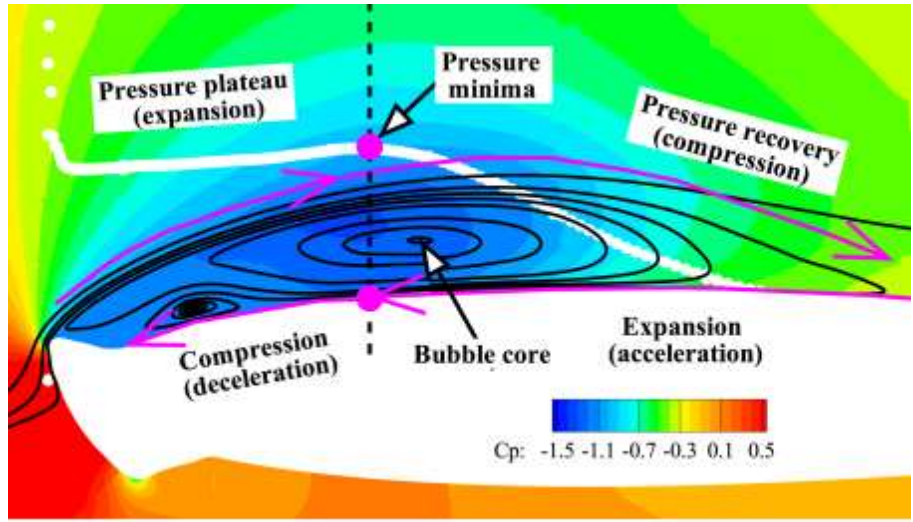

Figure 4 Characteristic time-averaged streamlines and pressure coefficients; white dots denote surface pressure coefficients (-$C_p$).

Figure 4 depicts the dominant separated flow characteristics at the midspan. The flow is dominated by a recirculation bubble bounded by the SSL emanating from the ice tip. Consequently, a surface pressure plateau forms around the separation bubble core and is then ensued by pressure recovery, which shares similar features with the surface pressure distribution within a laminar separation bubble [1, 26]. It is worth noting that the plateau actually exhibits a slow pressure decrease because the reverse velocity reaches the highest value exactly below the recirculation core. Figure 5 compares the calculated surface pressure with experimental measurements [24]. The results from the SST $k-\omega$ RANS are also shown, although they cannot reach convergence and show conspicuous errors. On the fine grid, both IDDES and IDDES-SLA produce almost the same results on the pressure side but show noticeable distinctions on the suction side. Specifically, IDDES-SLA is capable of correctly predicting pressure plateaus, whereas the flat pressure regions are elongated and lowered by IDDES, especially near the wing root, which directly leads to an excessive drag and nose-down pitching moment (Figure 3). The pressure distributions predicted via IDDES-SLA on the coarse grid almost coincide with those obtained on the fine grid. Hence, only the results obtained on the fine grid will be analysed hereafter.

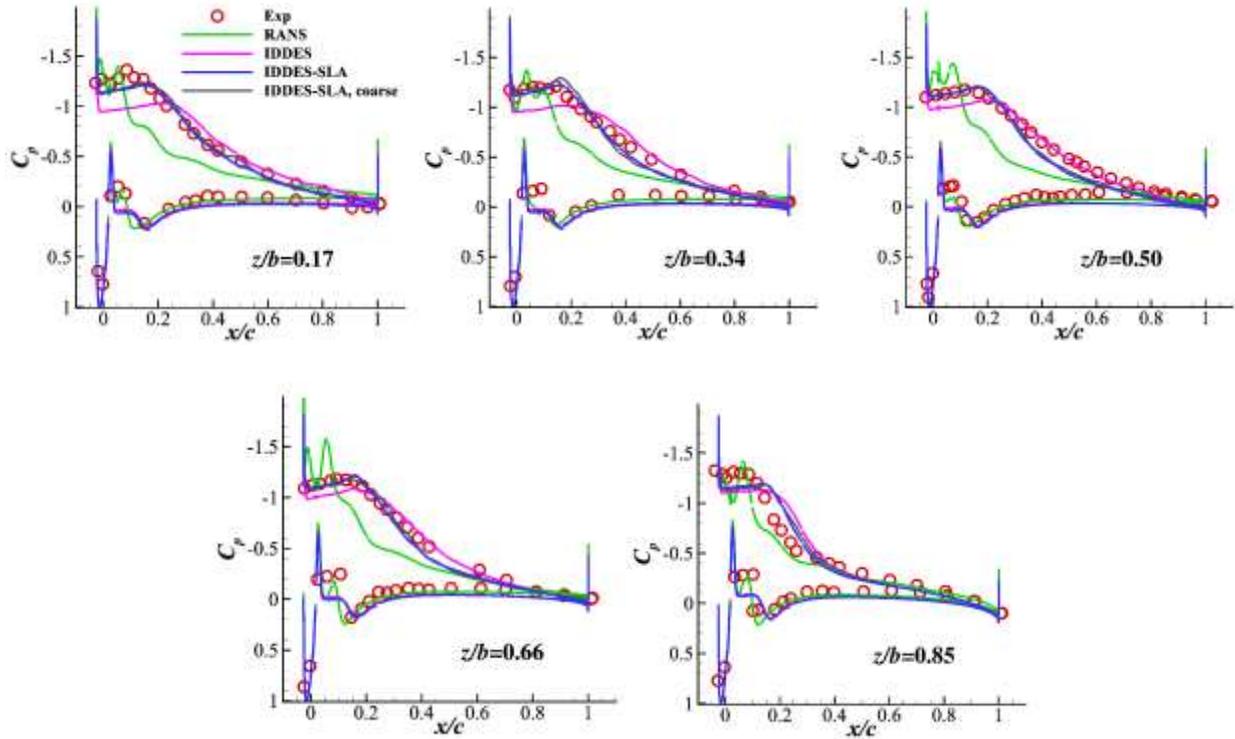

Figure 5 Time-averaged surface pressure coefficients. Experimental results [24].

Figure 6 further displays the time-averaged streamwise velocity fields superimposed by the streamlines for $z/b$=0.17, 0.50 and 0.85. The locations of reattachment $L_r$ and recirculation core $L_c$ are listed in Table 4. The relative difference denotes the percent by which an IDDES value is greater than its IDDES-SLA counterpart. Conspicuously, IDDES predicts lengthened recirculation regions in contrast to IDDES-SLA. At $z/b$=0.17, the IDDES-predicted recirculation is 57% longer than that of IDDES-SLA. The discrepancies are closely associated with the more downstream recirculation cores predicted by IDDES, which will be further discussed in section Ⅳ. B. In the experiment, the reattachment occurs more upstream near the wing root than at the midspan due to the end-wall effect [24]. It is captured by IDDES-SLA but not IDDES, which conversely extends the recirculation region when moving towards the wing root. In the wing tip region, the recirculation bubble is relatively small, which results from the downwash effect of the tip vortex.

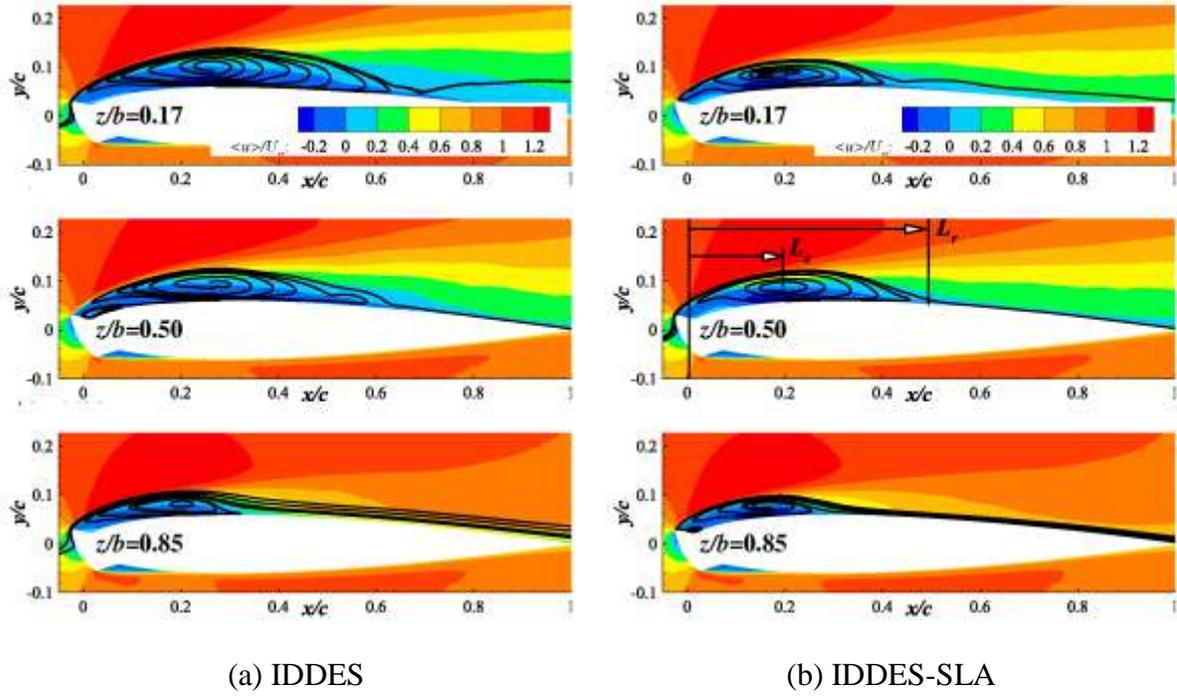

(a) IDDES  (b) IDDES-SLA

Figure 6 Time-averaged streamwise velocity superimposed by streamlines at different spanwise locations.

Table 4 Locations of the recirculation core and reattachment

| Method | $z/b$=0.17 | | $z/b$=0.50 | | $z/b$=0.85 | |
| --- | --- | --- | --- | --- | --- | --- |
| | $L_c/c$ | $L_r/c$ | $L_c/c$ | $L_r/c$ | $L_c/c$ | $L_r/c$ |
| IDDES | 0.26 | 0.69 | 0.28 | 0.64 | 0.20 | 0.30 |
| IDDES-SLA | 0.19 | 0.44 | 0.20 | 0.50 | 0.16 | 0.29 |
| Relative difference | 37% | 57% | 40% | 28% | 25% | 3% |

The discrepancies in the predicted recirculation regions make it crucial to examine the evolution of SSLs from different approaches. Here, the development of SSL is measured by the vorticity thickness:

$$\delta_\omega = \frac{\langle u \rangle_{max} - \langle u \rangle_{min}}{\max\left(\dfrac{\partial \langle u \rangle}{\partial n}\right)} \tag{8}$$

where $\langle u \rangle_{max}$ and $\langle u \rangle_{min}$ are the time-averaged maximum and minimum streamwise velocities, respectively, and $\partial \langle u \rangle / \partial n$ denotes the streamwise velocity gradient in the shear-normal direction. Figure 7 displays the streamwise evolution of the vorticity thickness for the midspan. The IDDES-SLA result includes three stages:

(1) $0<x/c<0.1$: the thickness grows exponentially, which is in accordance with the linear stability analysis of a classical mixing layer;

(2) $0.1<x/c<0.2$: the thickness growth shows a constant rate $d\delta_\omega/dx=0.32$. It is close to 0.33 from Pape et al. [27] for the SSL from an airfoil leading edge and 0.36 from Deck and Thorigny [28] for an axisymmetric separating-reattaching flow. These values are approximately twice the value of 0.17 for a classical mixing layer, which may be attributed to the turbulent fluctuations in the recirculation region [28]. In contrast, the growth pattern obtained via IDDES shows a much lower growth rate of 0.19;

(3) $0.2<x/c<0.5$: the shear layer thickness changes slowly due to reattachment, which is in accordance with the approximate plateau reported in [28]. However, the IDDES does not show small growth rates.

The ratio of $\delta_\omega$ to momentum thickness $\theta$ obtained via IDDES-SLA is also displayed in Figure 7. The momentum thickness is calculated as

$$\theta = \int_{y_{\min}}^{y_{\max}} \frac{\langle u \rangle - \langle u \rangle_{\min}}{\langle u \rangle_{\max} - \langle u \rangle_{\min}} \left(1 - \frac{\langle u \rangle - \langle u \rangle_{\min}}{\langle u \rangle_{\max} - \langle u \rangle_{\min}}\right) dy \tag{9}$$

The ratio is close to 5 for most of the SSL, which is consistent with other studies [28, 29].

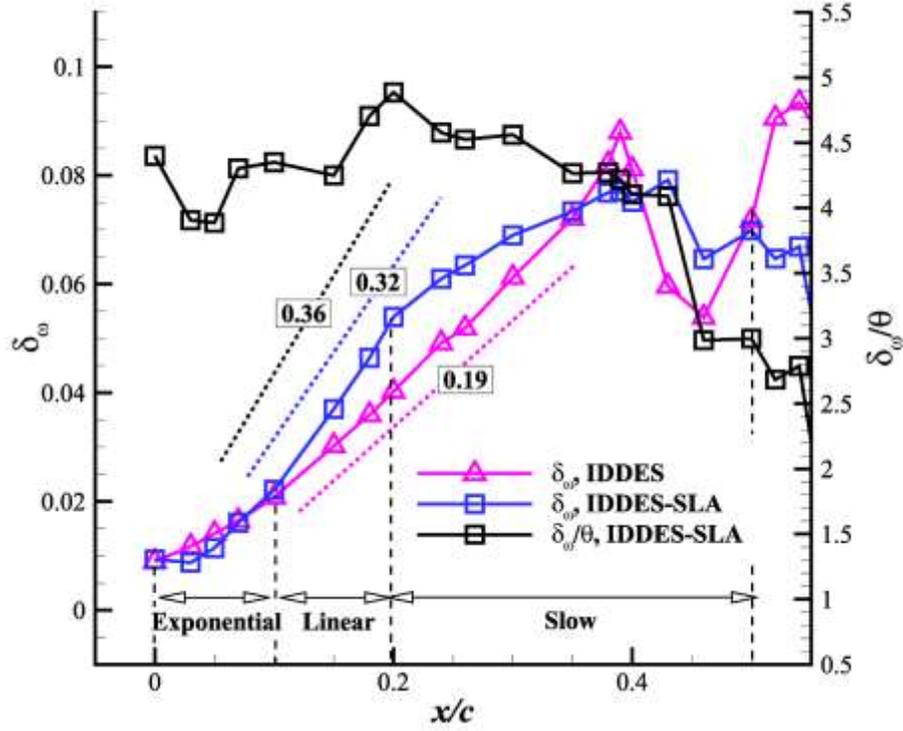

Figure 7 Vorticity thickness and the ratio of the vorticity thickness to the momentum thickness at the midspan.

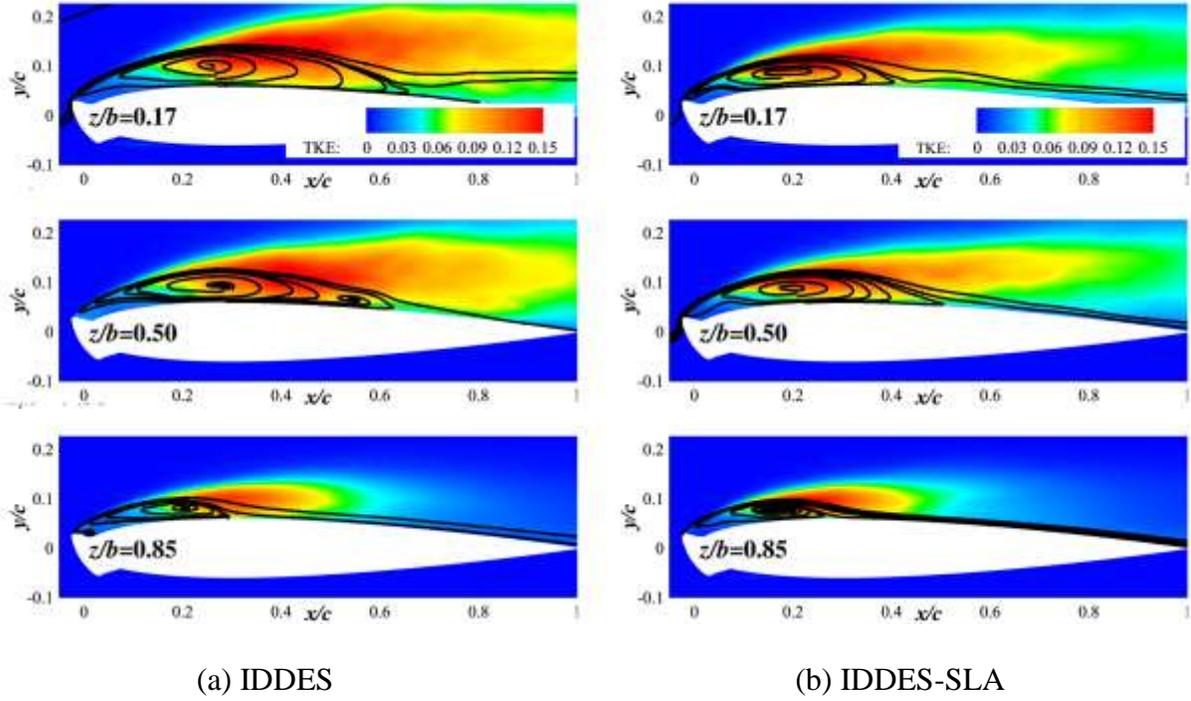

(a) IDDES  (b) IDDES-SLA

Figure 8 Time-averaged TKE at three spanwise locations.

The evolution of the SSL is closely determined by turbulent fluctuations that dominate fluid mixing. Figure 8 displays the time-averaged TKE obtained via the two methods. Obviously, IDDES predicts extended regions of intense fluctuations, which explains the higher RMS levels of integrated forces/moment (Table 3). However, it is the TKE from IDDES-SLA that grows faster in the initial SSL, which is crucial to correctly predicting the spreading rates (Figure 7) as well as the separation regions (Figure 6).

**B. Dominant Flow Structures**

The above analysis reveals strong unsteadiness along the SSL and downstream of the reattachment, which implies abundant unsteady flow structures. Figure 9 displays the flow structures identified by the $Q$ isosurface. The characteristics near the midspan are similar to those of a 2D iced airfoil [8]. The flow detaches from the ice tip, and 2D spanwise structures roll up due to K-H instability. Then, spanwise disturbances are gradually amplified due to a secondary instability, which promotes the formation of 3D structures, such as hairpin vortices. These vortices enhance the mixing of the inner low-speed reverse flow with the outer high-speed flow, which facilitates shear layer reattachment. Generally, the process has been resolved by both methods, but the IDDES-SLA prediction shows two superiorities:

(1) The initial 2D spanwise structures emerge rapidly and show small scales in the *x-y* plane, while IDDES produces delayed K-H instability.

(2) Secondary instability (spanwise destabilization) develops correctly, whereas it is delayed in IDDES, which results in "overcoherent" structures.

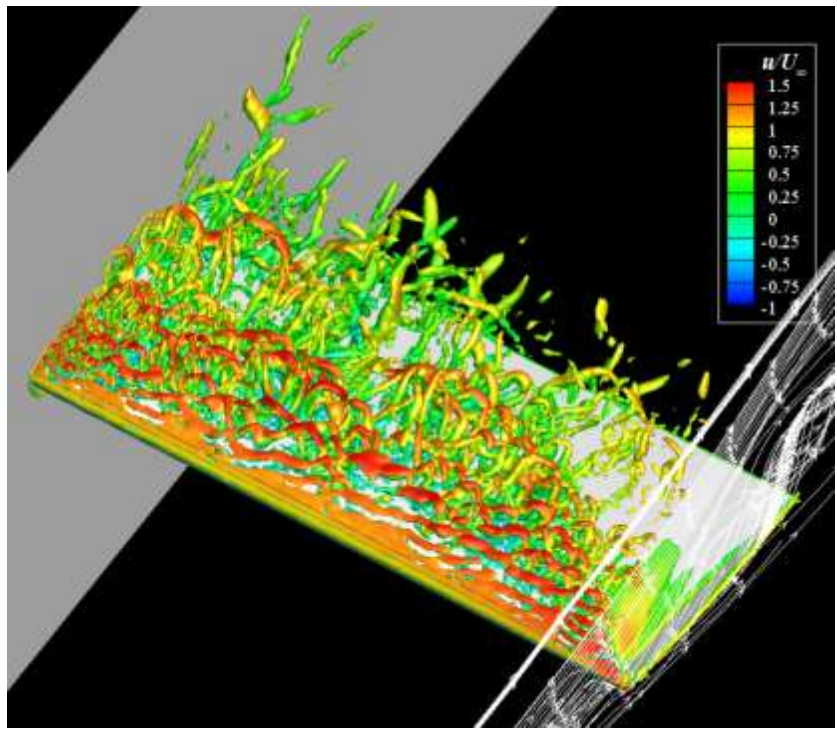

(a) IDDES

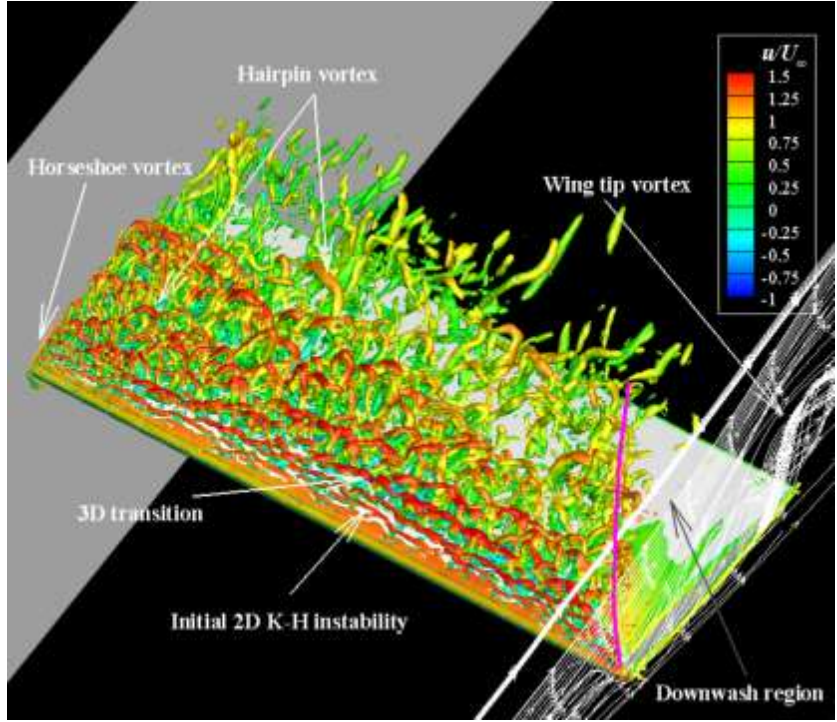

(b) IDDES-SLA

Figure 9 Instantaneous isosurface $Q(c/U_\infty)^2 = 100$ coloured by streamwise velocity $u/U_\infty$.

These advantages are visualized more clearly by Figure 10, which displays the instantaneous *y*-velocity at the midspan overlaid by the *Q* isoline. In contrast to IDDES, IDDES-SLA predicts rapid K-H instability both at the midspan and near the wing tip. To demonstrate the second advantage, Figure 11 compares the spatial-temporal variations of *y*-velocity at $(x, y, z)/c = (0.21, 0.11, 0.5-1.2)$, which is situated above the recirculation core (P6 in Figure 13). Both results exhibit quasiperiodic spanwise structures, but the structures obtained using IDDES exhibit larger scales in the spanwise direction. Additionally, they have longer time periods so that they have larger scales in the streamwise direction since the mean velocities from the two methods are very close. Deck and Thorigny [28] referred to the large-scale structures as "overcoherent" structures and indicated that they result in strong fluctuations near reattachment, which is consistent with the extended regions of strong fluctuations yielded from IDDES (Figure 8). In their axisymmetric SSL, the "overcoherent" structures are attributed to insufficient azimuthal mesh resolutions because increasing the resolution reduces the numerical errors. In the current study, the errors should be ascribed primarily to the modelling errors of IDDES since IDDES-SLA yields much better results. As shown in Figure 12, IDDES-SLA yields close-to-zero eddy viscosity due to the small hybrid length scales in the initial SSL. In contrast, IDDES generates excessive eddy viscosity that severely retards the occurrence of K-H instability and subsequent secondary instability, which finally leads to delayed spanwise structures and downstream "overcoherent" structures.

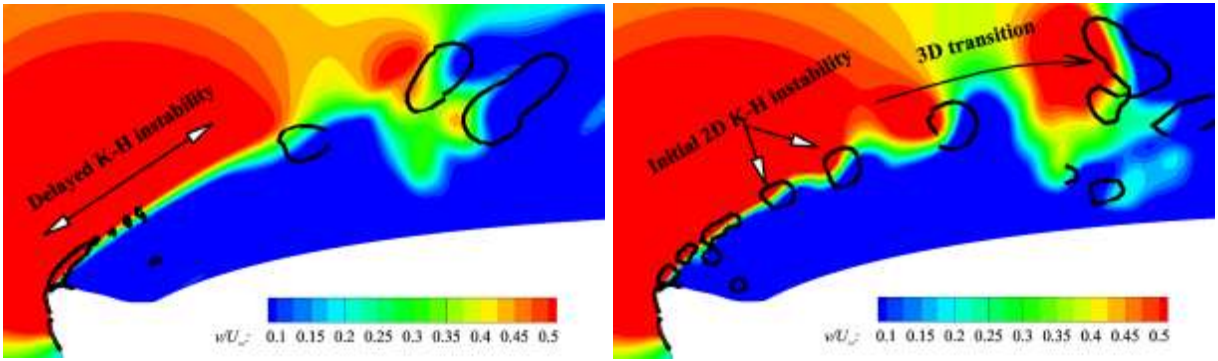

(a) IDDES, midspan  (b) IDDES-SLA, midspan

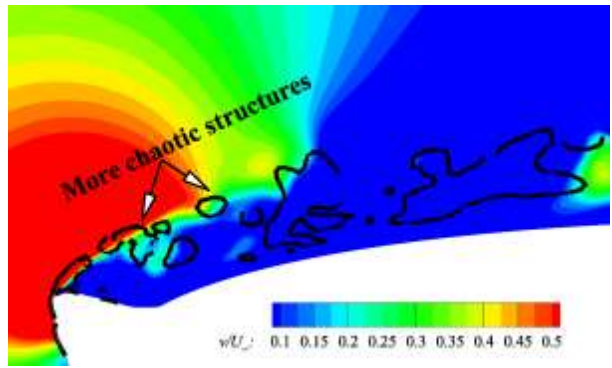

(c) IDDES-SLA, near the wing tip ($z/b$=0.92)

Figure 10 Instantaneous *y*-velocity component superimposed by isoline $Q(c/U_\infty)^2 = 800$.

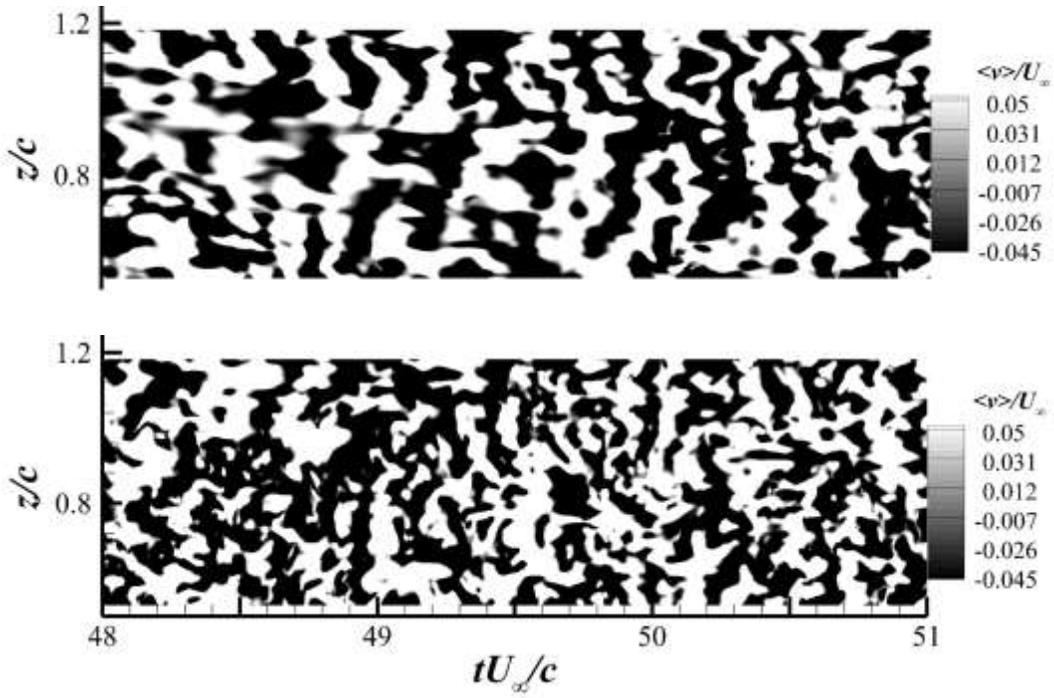

Figure 11 Spatial-temporal distributions of the *y*-velocity component at $(x, y, z)/c$ =(0.21, 0.11, 0.5-1.2). Top: IDDES; bottom: IDDES-SLA.

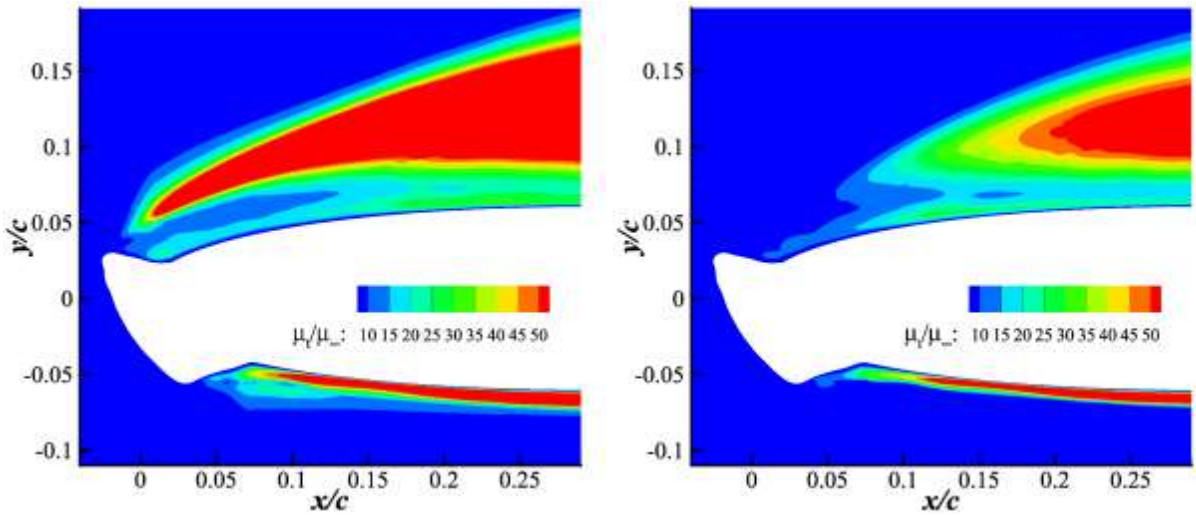

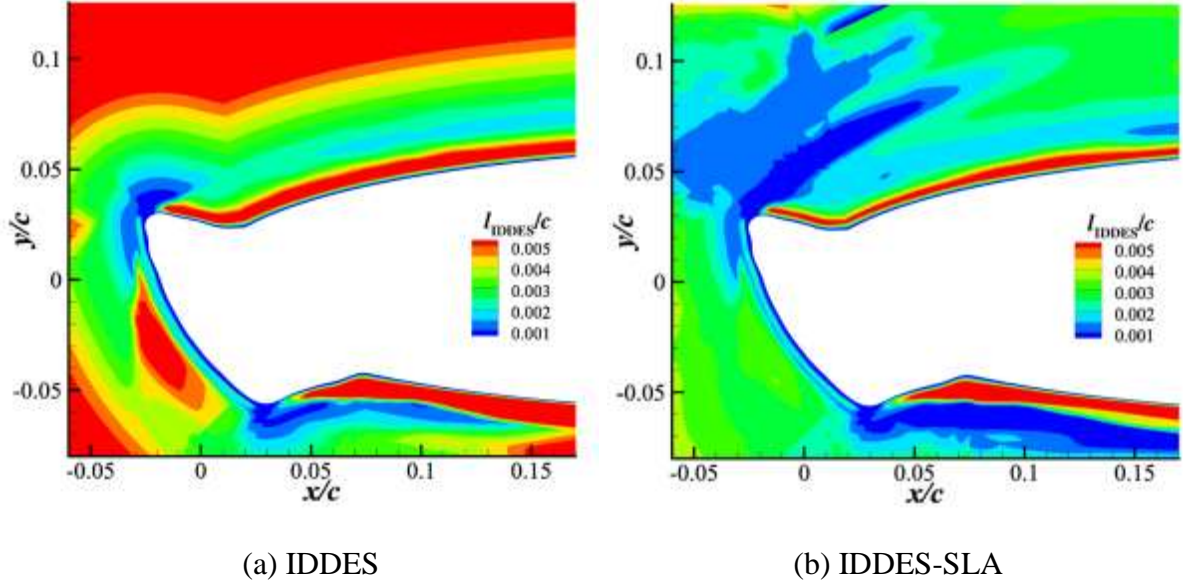

(a) IDDES            (b) IDDES-SLA

Figure 12 Time-averaged eddy viscosity (top) and hybrid length scale (bottom) at the midspan.

Moving towards the wing tip in Figure 9, the narrower regions are dominated by detached eddies, consistent with the earlier reattachment (Figure 6). This is due to the downwash effect of the tip vortex; the effective AoAs near the tip are lower than those at the midspan. In addition, the end-wall effects should be of interest since real aircraft experience this type of interaction in the wing-fuselage juncture. In Figure 9, a horseshoe vortex inhabits just upstream of the leading edge at the root section. This distorts the wing root flow along the turbulent boundary layer on the tunnel wall. This finally facilitates reattachment to shorten the recirculation length and reduce the sectional lift.

**C. Spectral analysis**

The power spectral density (PSD) is employed to analyse how unsteadiness is distributed in the frequency domain (refer to Figure 13 for the locations of sampling stations). Figure 14 (a) and (b) show the pressure spectra at stations P2, P3 and P5 for the midspan. It is observed that as one moves downstream, the low-frequency PSD levels gradually increase. Compared with IDDES, the spectra from IDDES-SLA show elevated PSD levels for each of the stations, which is expected since higher RMS levels are yielded from IDDES-SLA in the initial SSL (Figure 8). Table 5 lists the characteristic frequencies predicted by IDDES-SLA for P2 and P3. Each of the two stations has two characteristic frequencies, with a low frequency of approximately half of the high frequency. Similar results have been reported for 2D iced airfoils [8, 13, 15] and other separating-reattaching flows [30], and the frequency halving is attributed to vortex pairing. Table 5 also calculates the nondimensionalized peak frequencies based on the vorticity thickness and local mean streamwise velocity.

$$St_{\delta_\omega} = \frac{f\delta_\omega}{\langle u \rangle_{ave}} = \frac{f\delta_\omega}{\frac{\langle u \rangle_{high} + \langle u \rangle_{low}}{2}} \quad (10)$$

The $St_{\delta_\omega}$ values are 0.17 and 0.21 at P2 and P3, respectively. These frequencies are caused by K-H instability, since they are close to the theoretical value of 0.143 for a classical mixing layer [31]. Indeed, the current frequencies are somewhat higher, which is reasonable since the separating-reattaching flow differs from a canonical mixing layer due to reattachment. Richez et al. [32] obtained 0.13~0.20 for the SSL emanating from a wing leading edge. In contrast, the result at P2 obtained via IDDES shows three peaks at approximately $fc/U_\infty = 6.3$, 8.8 and 10.7, which does not show frequency halving. At P3, the spectrum seemingly exhibits two peaks at approximately $fc/U_\infty = 6.3$ and 4.3, but the low frequency is far from half of the high frequency. This may be explained by Figure 10, where the regular rollup of 2D vortices is reasonably captured by only IDDES-SLA. The frequencies from IDDES are not included in Table 5 since there are no theoretical $St_{\delta_\omega}$ values for ice-induced SSLs available for verification purposes.

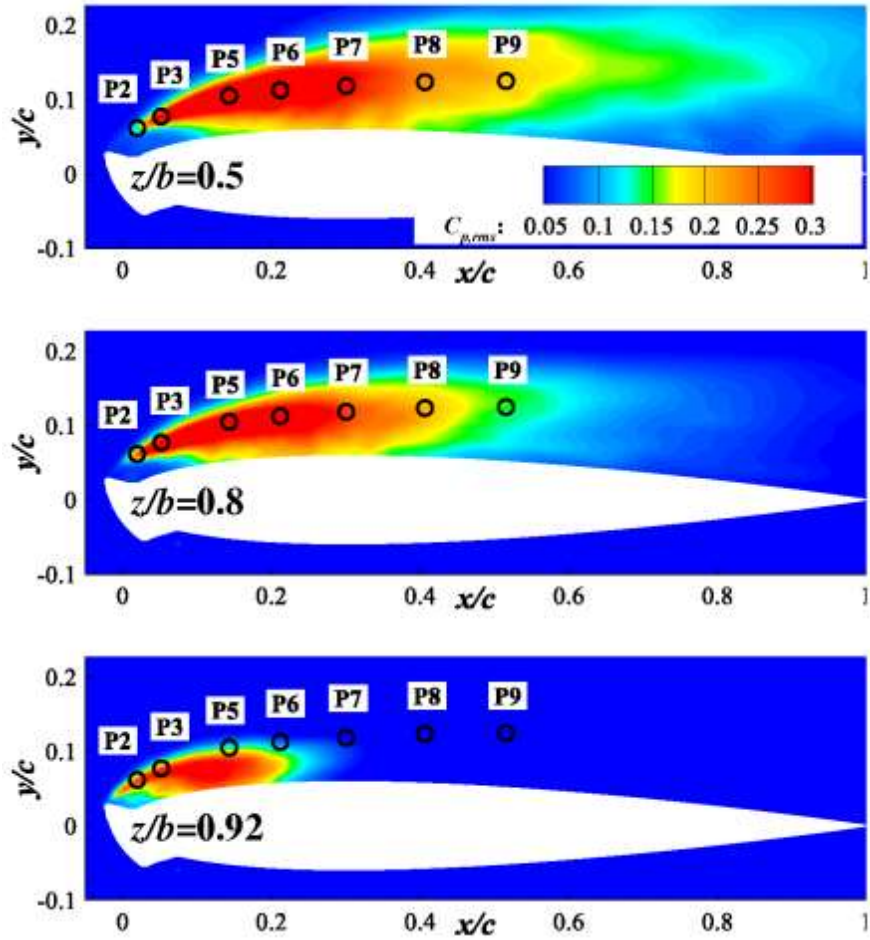

Figure 13 Sampling stations for spectral analysis; the contour is obtained via IDDES-SLA.

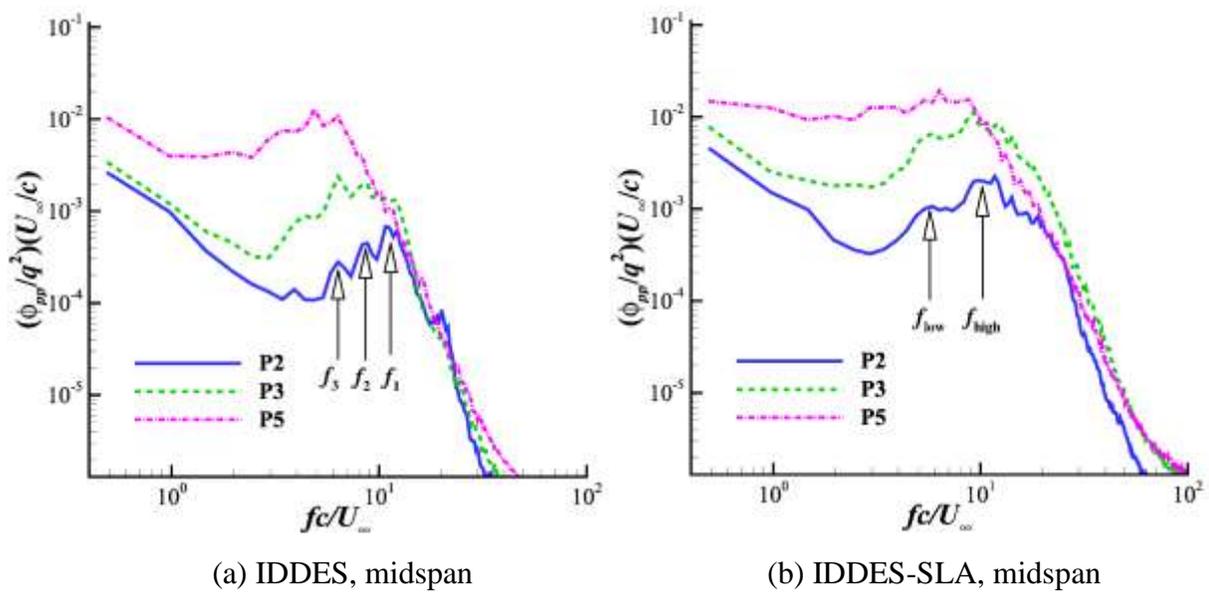

(a) IDDES, midspan  (b) IDDES-SLA, midspan

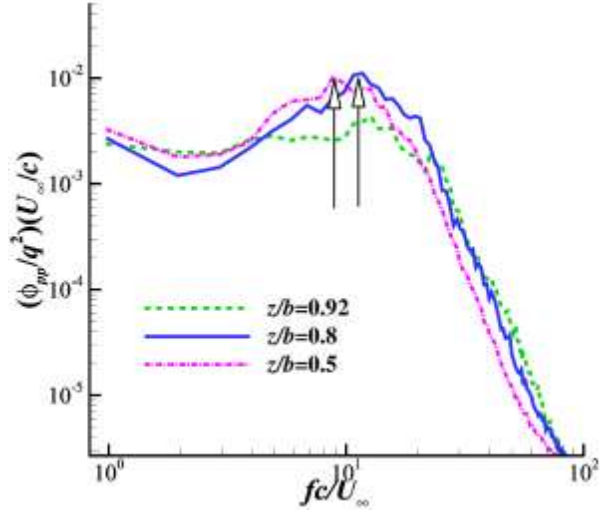

(c) IDDES-SLA, P3, wing tip

Figure 14 Spectra of pressure fluctuations in the initial SSL.

Table 5 Characteristic frequencies in the initial SSL from IDDES-SLA

| Station | $\delta_\omega$ | $\langle u \rangle_{ave}$ | $f_{high}$ | $f_{low}$ | $St_{\delta_\omega, high}$ |
|---------|-----------------|---------------------------|------------|-----------|----------------------------|
| P2      | 0.0089          | 0.57                      | 10.8       | 5.6       | 0.17                       |
| P3      | 0.0127          | 0.57                      | 9.2        | 5.4       | 0.21                       |

The improved prediction of K-H instability by IDDES-SLA at the midspan provides confidence for examining the SSL near the wing tip. Figure 14 (c) displays the spectra at P3 for $z/b$=0.80 and 0.92. To obtain smooth curves, the spectrum at each spanwise location is an average of the spectra at 20 neighbouring stations with the same ($x, y$) but different $z$. Specifically, the station $z/b$=0.80 is obtained by averaging the stations between $z/b$=0.76-0.83, and the station $z/b$=0.92 is the average of stations between $z/b$=0.89-0.94. The spectrum at $z/b$=0.5 is also shown for comparison. It seems that the spectrum at $z/b$=0.8 can be obtained by shifting the spectrum at $z/b$=0.5 to the right. Obviously, the spectrum at $z/b$=0.8 shows a peak at approximately $fc/U_\infty = 11.2$, higher than $fc/U_\infty = 9.2$ for the midspan. The distinction can be clearly visualized by the spatial-temporal y-velocity variations shown in Figure 15. The average time period of vortical motions is shorter at $z/b=0.8$ than at the midspan, since more spanwise structures exist within the same time interval. In a previous work on a 2D iced airfoil [13], the vortical motion frequencies in the initial SSL are increased by approximately 10% when the AoA is decreased by 1.8 deg, which may be associated with the increased fluctuations below the initial SSL. Indeed, Figure 13 shows increased fluctuations below the initial SSL towards the wing tip. Therefore, the increased frequency near the tip is associated with the downwash effect of the tip vortex. Moving to $z/b=0.92$, the peak becomes less conspicuous, which may be correlated to the more chaotic structures compared with those at the midspan (Figure 10 (c)).

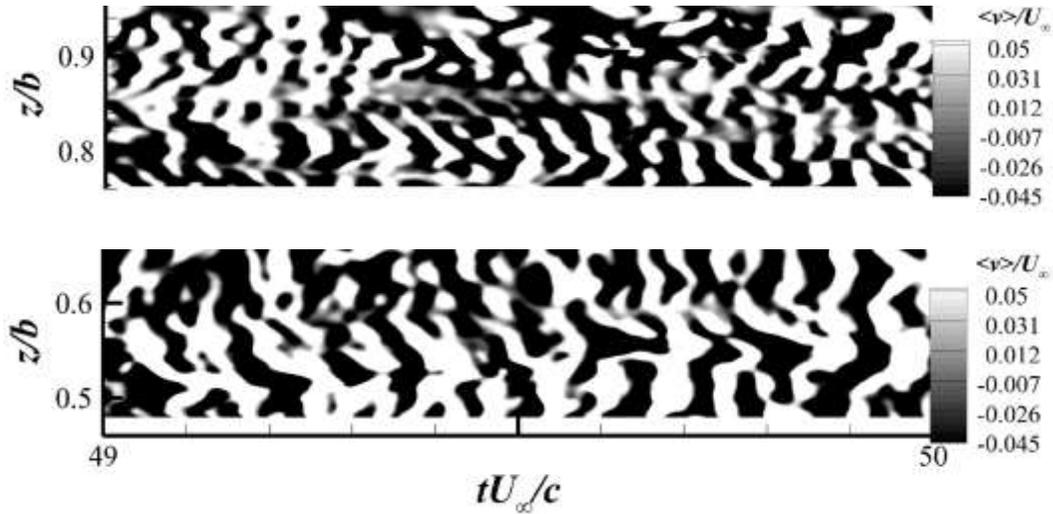

Figure 15 Spatial-temporal distributions of y-velocity at P3. Top: wing tip; bottom: midspan.

Moving further downstream, Figure 16 (a) shows the spectra near reattachment at P7, P8 and P9 for the midspan. Each of the spectra exhibits a pattern that follows an $f^{-2}$ dependency between $fc/U_\infty = 2\sim10$, which is consistent with Terraco and Manoha [33] for the SSL emanating from a slat cusp. From P7 to P9, the PSD levels gradually decrease due to the decreasing levels of pressure fluctuations (Figure 13). The spectra do not show conspicuous dominant frequencies, but the temporal variations in pressure do exhibit quasiperiodic patterns, as illustrated by the inserted plot in Figure 16 (a). It labels two representative periods $TU_\infty/c$ = 0.42 and 0.48, so that the corresponding frequencies are $fc/U_\infty$ = 2.34 and 2.08. The frequencies are associated with the vortex shedding near reattachment, since the nondimensionalized frequencies based on the recirculation length are $fL_r/U_\infty$ =1.1 and 1.0, close to the frequency range of 0.48-1.0 for various flows [34]. The frequency $fc/U_\infty$ = 2.08 is arrowed in Figure 16 (a), and it approximately corresponds to the turning points of the spectra. The interpretation of the vortex shedding is not unanimous, but most studies support its relation with the vortical motions in the SSL. For instance, vortical structures from an SSL agglomerate near the reattachment before being advected downstream, which causes the instantaneous reattachment location to move in a quasi-periodic manner [28]. Figure 16 (b) further displays the pressure spectra near the reattachment for $z/b$= 0.8 and 0.92. The spectra show decreased PSD levels in the low frequency range in contrast to that at the midspan. If we estimate the vortex shedding frequency as the turning point of the spectrum, it becomes higher when moving towards the wing tip and thus results in a nearly constant $St_{L_r}$ since the recirculation length becomes shorter.

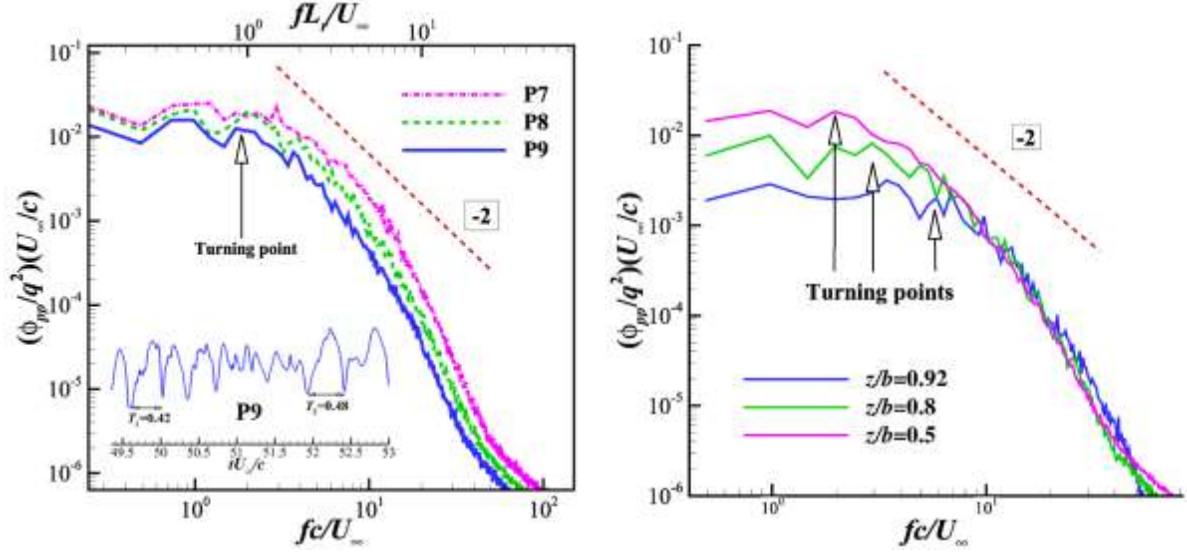

(a) Pressure spectra, midspan        (b) Pressure spectra, wing tip

Figure 16 Spectra of pressure fluctuations near reattachment.

## V. Conclusions

A shear layer adapted subgrid length scale is employed to improve the prediction of the separating-reattaching flow around a 3D NACA 0012 wing with leading edge horn ice. Its superiorities are demonstrated via comparison with the original IDDES in terms of instantaneous and statistical flow data and spectral characteristics. Additionally, the effects of the tip vortex and end wall are investigated.

IDDES-SLA is capable of correctly predicting the sectional lift, surface pressure and surface flow. The predicted spanwise variation of the separation region agrees well with the experiment. Specifically, it becomes shorter either as one approaches the wing tip due to the downwash effect of the tip vortex or as one moves towards the wing root due to the end-wall effect. The resulting shear layer growth exhibits three stages, and the linear growth part shows a nearly constant spreading rate $d\delta_{\omega}/dx = 0.32$, which is close to 0.36 from other studies [28, 32]. In contrast,

IDDES predictions show large discrepancies in the drag and pitching moment because they produce extended surface pressure plateaus and separation regions. Additionally, IDDES yields extended separation regions from the midspan to the root, which is opposite to the trend in the experiment.

Both the vortex rollup and pairing have been correctly captured by IDDES-SLA due to its low eddy viscosity levels resulting from its small hybrid length scales in the initial SSL. The nondimensionalized frequencies $St_{\delta_\omega}$ of vortical motions are 0.17-0.21, which are in accordance with 0.13-0.24 in various SSLs [28, 32]. In addition, IDDES-SLA shows a nondimensionalized frequency $St_{L_r}$ ~1.0 near the reattachment, consistent with 0.48-1.0 for the vortex shedding mode in different separating-reattaching flows [34]. Both the K-H instability and vortex shedding modes exhibit higher frequencies when moving towards the wing tip, which is correlated with the downwash effect of the tip vortex. In contrast, the excessive eddy viscosity of IDDES results in the delayed rollup of spanwise structures and the formation of "overcoherent" structures; vortex pairing is not reasonably captured either.

# Acknowledgements

The work was supported by the National Natural Science Foundation of China (Nos. 11872230 and 92052203) and the Aeronautical Science Foundation of China (No. 2020 Z006058002).

# Data availability

The experimental data can be found in the references [24, 25]. The computational data that support the findings of this study are available from the corresponding author upon reasonable request.